# Extrinsic plastic hardening of polymer thin films in flat punch indentation

Owen Brazil[1,2], Johann P. de Silva[1], John B. Pethica[1], Graham L. W. Cross[1,3]*

[1]*School of Physics, CRANN & AMBER, Trinity College Dublin, Ireland*

[2]*Department of Materials Science & Engineering, Texas A&M University, College Station, TX 77840, USA*

[3]*Adama Innovations LLC, CRANN, Trinity College, Dublin 2, Ireland*

*Correspondence to: crossg@tcd.ie

# Abstract

Confined geometries offer useful and experimentally amenable mechanical testing arrangements in which to study the molecular and micro-structural processes which govern plastic yield in stress environments dominated by hydrostatic pressure over shear. However, the changes to macroscopic stress – strain behaviour that result from switching from an unconfined mode such as uniaxial compression to a confined one are often overlooked and display a surprising level of complexity, even for simple elastic – plastic constitutive models. Here we report a confinement induced strain hardening effect in polystyrene thin films achieved through repeated plastic loading with a cylindrical flat punch whose diameter is many times the initial film thickness. This high aspect ratio combines with constraint provided by film material surrounding the contact to generate a state of confined uniaxial strain in the indented region, rendering the deformation one dimensional. By repeated loading past the yield point into the plastic domain, we achieve a 66% increase in the confined yield stress, from 0.3 GPa to 0.5 GPa. Through finite element simulation and analytic modelling of the principal stresses and strains, we show that this effect arises not from intrinsic changes to the structure of the material, but rather residual stresses imparted during plastic loading. We contrast this effect with intrinsic changes to glassy thin films such as physical ageing and thermal cross-linking.

# Main Text

## 1. Introduction

Materials confined in one or more direction often exhibit mechanical behaviours that differ significantly from their response in unconstrained deformation modes like simple shear or uniaxial tension. This is particularly true in the plastic limit where the microstructural mechanisms underpinning deformation may have non-intuitive dependencies on pressure or shear. For example, bulk metallic glasses deformed in deeply notched geometries that effectively render material displacement one-dimensional have shown a densification and strain hardening effect associated with a preferential free volume annihilation mechanism not seen in standard tensile testing, as well as a suppression of shear banding.[1], [2] Confinement effects are particularly relevant at small scales, where a sample length scale, e.g. film thickness, is reduced to the point where defect or molecular rearrangement in one or more directions becomes unfavourable, leading to a divergence from bulk properties[3]–[6]. Examples are numerous and include reduced viscosity for polymer films in squeezed nanoimprint geometries[7], [8] and the enhanced strength and toughness of nanolaminate materials.[9], [10] Broadly speaking, the majority of research has focused on the influence of confinement on the molecular/microstructural level processes responsible for elastic and plastic deformation of the material.[11] Borrowing from the vocabulary of small scale mechanical testing, these can be considered intrinsic confinement effects,[12] with extrinsic effects being those that would be experienced by an idealised continuum material in a similar geometry with no internal length scale such as radius of gyration or Burger's vector. Deconvolution of these intrinsic and extrinsic effects is critical in correctly interpreting the influence of confinement on material response, however, often proves challenging, as to correctly characterise the latter requires that the stresses and strains throughout the material be well known.[10] This is far from trivial for complex deformation geometries and is further complicated in the presence of significant plastic deformation, where explicit relations between stress and strain generally do not exist even for the simplest constitutive models and incremental flow rules must be employed.[13] The classic example of an extrinsic confinement effect is Tabor's constraint factor $H = CY_0$, which relates indentation hardness $H$ to tensile yield stress $Y_0$ for conical and pyramidal indenters.[14] During indention of an elastic-plastic half-space, first yield occurs at some depth below the surface and as such is constrained by the surrounding elastic material. Due to the complexity of the deformation however, no general solution exists for the stresses within the half-space and simplified models such as Johnson's expanding cavity must be employed.[15] This leads to a constraint term $C$ that depends not only on the material parameters and indenter geometry, but also upon the model employed.[16]

In contrast, we recently have shown that a nearly uniform state of confined compression uniaxial strain (CC) can be achieved in glassy thin polymer films through an indentation technique we call the layer compression test.[17] Geometric confinement allows isolation and controlled study of distinct elastic and elastic-plastic states at the nanoscale with sensitive, low stiffness instruments. Prior to this, CC had only been demonstrated in granular or porous materials due to difficulties in preventing Poisson driven expansion and/or volume preserving plastic flow.[18], [19] The ideal CC test geometry is sketched in Fig. 1a. A compressive stress $\sigma_{zz}$ is applied to a sample encased by a rigid jacket which prevents lateral strain. This results in a stiffer elastic response than conventional unconfined compression, and the development of significant radial stresses $\sigma_{rr}$ on the jacket. Absence of lateral frictional traction means deformation is highly uniform throughout the sample with simple, explicit relations between all principal stresses and strains. In the elastic regime, axial (z), radial (r) and azimuthal (*θ*) applied stress $\sigma$ is related to resultant strain $\varepsilon$ as:

$$\sigma_{zz} = E\frac{(1-v)}{(1+v)(1-2v)}\varepsilon_{zz} = M\varepsilon_{zz}, \quad \sigma_{rr} = \left(\frac{v}{1-v}\right)\sigma_{zz}, \quad \varepsilon_{rr} = \varepsilon_{\theta\theta} = 0 \quad (1.a,b,c)$$

where $E$ is Young's modulus, $v$ is Poisson's ratio, and $M$ is referred to as the confined elastic modulus and is larger than $E$ for $v$ positive.[20] For simple von Mises plasticity, yield occurs at an elevated confined yield stress:

$$Y_c = (1-v)/(1-2v)\, Y_0 \quad (2)$$

where $Y_0$ is the unconfined tensile/compressive yield stress. Beyond the point of yield, stable, homogenous one-dimensional plastic compression in the vertical direction is maintained by lateral confinement, a unique feature of CC mechanics. In stark contrast to other deformation modes, explicit relations for stress vs. strain exist in the plastic state for materials that strain soften or have little work hardening. For an elastic perfectly plastic material loaded past the yield point, the stress-strain relation becomes $\sigma_{zz} = K\varepsilon_{zz}$, where $K$ is the elastic bulk modulus.[21]

The layer compression test [17] realizes CC by indentation of an aligned, cylindrical flat punch whose diameter is many times the initial film thickness (Fig. 1b). The large punch diameter to film thickness aspect ratio $\alpha = 2a/h_0$ combines with confinement afforded by film material surrounding the contact to suppress lateral strains and render the deformation uniaxial. Under an applied load, $L$, the mean axial stress is readily obtainable as $\sigma_{zz} = L/\pi a^2$ while the engineering strain is obtained from the current film thickness $h$ as $\varepsilon_{zz} = 1 - h/h_0$. Deformation uniformity implies these mean quantities are close to the local stress and strain experienced material anywhere under the punch. Fig 1c. presents the stress-strain response of a $h_0$ = 240 nm atactic polystyrene (aPS) film bonded to a silicon substrate indented with a *2a =* 2050 nm diamond punch at a constant stress rate of 0.67 GPa/s. Linear fits to the confined elastic and confined plastic regions of the curve are shown in blue, with the associated moduli

*M* and *K* identified as well as the confined yield stress $Y_c = 0.32$ GPa. At a stress of $\sigma_{zz} = Y_{flow} = 0.65$ GPa, the confining material yields leading to extrusional flow and gross deviation from the CC state. This confinement failure value has been shown to depend on the geometric and material parameters of the test.[17] Until $Y_{flow}$, the deformation is highly one dimensional as demonstrated in Fig 1d & 1e, which show stress – strain curves and AFM residual topography maps for three separate indents of a $h_0 = 190$ nm aPS film. For the first indent to $\sigma_{zz} = 0.28$ GPa (blue), the response is elastic, with no residual strain evident either in the stress-strain curve upon unloading or in the AFM image of the film. The $\sigma_{zz} = 0.5$ GPa (pink) indent generates ~10 nm of residual strain upon unload. Examination of the AFM image indicates that this manifests as a circular crater in the former position of the punch. Pile-up is not observed in the hinterland surrounding the crater, indicating that a 1D densification process has occurred upon yield, consistent with predictions of CC for an elastic plastic material. In the yellow curve of Fig 1d which loads to a stress above $Y_{flow}$, a halo of extruded pile-up material surrounding the contact is observable in Fig 1e.

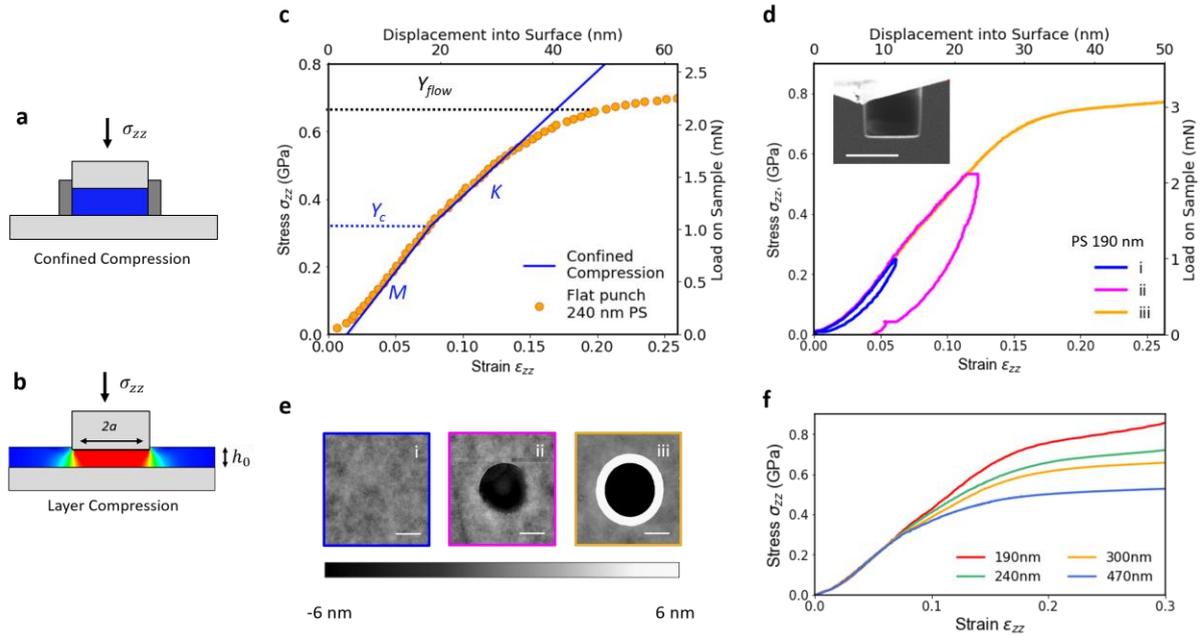

**Figure 1. Flat punch indentation approximates uniaxial confined compression in thin films.** 2D schematic of pure CC (a) vs. layer compression via flat punch (b). (c) Load vs. displacement converted to stress – strain curve for 240 nm aPS film indented by 2050 nm diameter punch with confined yield at 0.32 GPa and linear fits (blue) to confined elastic and plastic regions. (d) Stress-strain curves for elastic, confined plastic and extruded regime peak loads on 190nm aPS film (e) AFM topography maps of the indents. (f) Scaling of stress – strain curve for PS with increase punch diameter to film thickness ratio.

The stress-strain behaviour described above is essentially an extrinsic confinement effect; constraint provided by film material surrounding the contract drives the system towards the unusual uniaxial strain state. However, without knowledge of uniaxial strain mechanics, many characteristics of the layer compression test may be misinterpreted as intrinsic effects. The heightened yield stress could be mistaken for a scale dependent strengthening of the polymer analogous to the Hall-Petch effect in crystalline metals.[22], [23] The confined plastic modulus $K$ also increases for thinner films as is shown in Fig 1f, however this is in fact due to increased deformation uniformity as the test aspect ratio is increased. Pancake geometries resembling the layer compression test are encountered throughout advanced materials testing but are generally analysed without consideration of the CC mechanical framework we provide here. For example, micropillar compression tests performed on a nanolaminate composite material consisting of alternating layers of amorphous $Cu_{54}Zr_{46}$ and crystalline Cu showed significantly greater ductility than monolithic pillars of either material, with strains greater than 40% achieved before fracture.[9] A mechanism of shear band propagation suppression was proposed to explain this behaviour; however it was not recognized that by virtue of the sample geometry each pancake-like nanolaminate layer was in a state of CC as opposed to conventional unconfined pillar compression. Caruthers et al. performed tensile uniaxial strain tests on a high aspect ratio epoxy sample close to its glass transition temperature $T_g$ in order to probe the molecular processes behind polymer yield.[24], [25] While equations (1 a-c) & (2) provided here were used to study the elastic behaviour and first yield, no analysis of stress and strain in the yielding state (e.g. Mises-Levy, Prandtl-Reuss) was performed and no relation between $\sigma_{zz}$ and $\varepsilon_{zz}$ identified. The change in slope at yield was therefore interpreted as an increase in segmental mobility equivalent to raising the temperature of the sample by several 10's degrees, as opposed to an extrinsic constraint effect.

In this study we further explore the implications of CC for simple materials via the layer compression test. Using glassy PS films at a temperature far below $T_g$ as an example material, we report a pseudo strain hardening effect upon repeated cycled indentation into the confined plastic limit, where the confined yield stress is observed to increase by as much as 66% from its initial value. Finite element simulations of a simple linear elastic-plastic material in pure CC show the same behaviour, indicating that the phenomenon is extrinsic in origin. Analysis of the principal stresses and strains during indentation as well as hydrostatic and deviatoric components reveal this effect originates from uniform residual radial stresses imparted to the sample during confined plastic deformation. We contrast this behaviour with increases in yield stress brought about through physical ageing in PS and thermal crosslinking in a nanoimprint micro-resist material. We believe this phenomenon to be of particular significance given the recent flurry of high-quality studies performed on plastically deformed bulk metallic glasses in similar deep notch test geometries, where to the best of our knowledge no analysis of residual stresses has yet been performed.[1], [2], [26]–[28]

## 2. Experimental Methods

### 2.1 Nanoindentation Experiments

Atactic polystyrene films (Polymer Source Inc.) of 1.13 MDa molecular weight and polydispersity ≈ 1 and thicknesses of 470 & 670 nm were prepared via spin coating from polystyrene/toluene solutions (1-3% wt. polystyrene) on silicon <100> wafer pieces of approximately 1 cm$^2$ (University Wafer.) Samples were then heated to $T_g$ + 30°C (130°C) for 30 minutes on a hotplate (Torrey Pines Scientific) to remove residual solvent content, before being stored under vacuum overnight. Film thickness was measured via profilometry using a Veeco Dektak 6M profileometer with sub nm accuracy and confirmed via AFM. Two diamond flat punches of diameters $2a$ = 2050 & 4800 nm were used in this work. Both were fabricated via focused ion beam milling of conventional cube corner nanoindenter tips (Micro-Star Technologies) using an FEI Strata 235 dual beam FIB-SEM system. Thin film samples were mounted to a dual axis tilt stage (Physik Instrumente M-044) capable of *μrad* precision via crystal bond (Electron Microscopy Sciences). Indents were preformed using an MTS Nanoindenter XP system. For the smaller punch the high-sensitivity dynamic contact module head was used, whereas the conventional head was used for the larger punch. Alignment was ensured between punch and sample using the tilt stage in tandem with a DME DS 95 AFM as has been described previously.  Indents were performed at a stress rate of 0.20 GPa/s during loading and unloading for cyclic loading experiments. A 1 nm vertical oscillatory displacement was applied to the punch at 45 Hz throughout.

**2.2 Thermal ageing of polystyrene and curing of thermoset films.**

In order to demonstrate the influence of intrinsic structural changes on the stress – strain response of polystyrene in CC a 550 nm PS film on silicon was prepared as in section 2.1.The sample was then loaded via high temperature cement (Omega CC) to a special mounting piece which could be inserted and removed from the indenter with minimal (< 0.05°) change in sample – punch alignment. For the quenched thermal history, the sample was heated to 130°C, held at this temperature for 30 mins, and then rapidly cooled to room temperature. The sample was then indented at a stress rate of 0.05 GPa/s. For the annealed history, the sample was again heated to 130°C and held there for 30 mins, before being cooled at a rate of 8°/hr to 75°C, where it was held for 5 hours.

To measure the effect of chemical crosslinking on the CC stress – strain curve, a 300 nm film of the commercially available micro-resist mr-I 9000 (Micro Resist Technology) was prepared. This material is based on a partially polymerized poly-diallyl phthalate (PDAP) pre-polymer, whose remaining allyl double bonds may react to generate a highly cross-linked network at temperatures greater than 120°C. [29], [30]. Following spin coating, the film was heated at 60°C in order to remove residual solvent. The sample was loaded onto the same mounting piece used for the thermal ageing experiments and indented before and after curing for 30 mins at 150°C.

**2.3 Finite Element Simulations**

Axisymmetric simulations were performed using the Abaqus 2019 Explicit (Dassault Systemes) finite element package. A pure confined compression uniaxial strain geometry was constructed of a $r = h_0 = 1\ \mu m$ cylindrical elastic-plastic material subjected to boundary conditions prohibiting lateral expansion. The punch was modelled as a perfectly rigid body of 1 μm radius with sharp corners. Material properties were chosen to approximate those of a glassy polymer far from $T_g$, namely: $E$ = 3.0 GPa, $v$ = 0.33, $Y_0$ = 0.1 GPa, and $\rho$ = 1.04 g/cm$^3$.[31] These values give a confined yield stress $Y_c$ of 0.203 GPa. 4-node bilinear axisymmetric quadrilateral (CAX4R) elements were utilized, with the initial element area set as 50 x 50 nm. Compression was simulated via prescribing a vertical displacement at a reference point attached to the punch face. No other displacements of the punch are allowed. A full-slip condition was specified between punch surface and sample. An encastre condition was applied to the bottom surface of the sample such that $U_r = \underline{U_z} = 0$, simulating a full-stick condition on a rigid substrate. A boundary condition setting radial displacements to zero was imposed at the sample wall, while vertical displacements were allowed with no friction. Principal stresses and equivalent shear and pressure stresses were extracted as field variables from the deformed material and compared to reaction forces on the punch reference point and on the sides of the sample.

## 3. Results and Discussion

In order to probe the plastically yielding CC state in greater detail than previously achieved, two sets of cycled indentation experiments were performed, shown in figure 2. The first protocol, plotted in Fig. 2a, consists of cycled loading to a single peak stress greater than the confined yield stress $Y_c$. The stress – strain response for a $h_0$ = 470 nm PS film indented with the 2050 nm punch ($\alpha$ = 4.4) is shown in Fig. 2b, where superposed on a simple monotonic loading curve (blue), the cycled stress vs. strain curve (red) shows no repeat of the initial yield kink at $Y_c$ on subsequent loading cycles. This indicates that loading in confined compression to beyond yield on the first cycle results in a permanently raised yield stress. A second cycling protocol shown in Fig. 2d performed with a 4800 nm punch on 670 nm PS film ($\alpha = 7.1$) to increasing peak stress $\sigma_1 < \sigma_2 < \sigma_3$ confirms this: yield occurs at increasing stress $Y_c < Y_{c2} < Y_{c3}$ in the stress – strain curves plotted in Fig. 2e. The maximum measured yield stress on the third cycle reaches a value of $Y_{c3} = 0.50$ GPa, approximately 66% higher than the initial $Y_c = 0.32$ GPa.

In glassy polymers, non-linear effects such as anelasticity, strain softening, strain hardening, and creep behaviour have led to an extensive collection of sophisticated continuum models to describe plastic deformation. Both experiment[32] and molecular dynamics simulations[33], [34] indicate that shear strain may underlie this, leading to accelerated aging due to exponential decreases in segmental relaxation times and allowing the material to evolve to lower local minima on a potential energy landscape. While remaining an open topic of current research, a leading mechanism thought to underpin plasticity in amorphous systems is the shear transformation zone,[35] where plastic deformation consists of localised, permanent rearrangements involving a small number of molecules which in turn exert elastic forces on the surrounding medium.[36] Such shear transformations have been observed in both atomistic simulations[37] and in experimental studies of colloidal glasses[38] and are mechanistically distinct from dislocation-based plasticity models. Nevertheless, we find that yield stress behaviour in CC is describable as an effect of the confined geometry using simple elastic-plastic constitutive relations originally developed for metals, e.g. a Von Mises distortion strain energy based yield criterion. Figs. 2c and 2f show equivalent FEA simulations of the two protocols discussed above for an isotropic linear elastic – plastic material in a pure CC state. The results are essentially identical to the layer compression test results, with Fig. 2c showing no subsequent yield after first loading to a single stress above $Y_c$, and Fig. 2f exhibiting the same characteristic increase in $Y_c$ upon each reloading cycle, albeit without the viscoelastic hysteresis seen in experiment. As this behaviour may be reproduced by simple continuum based simulations, we conclude it cannot be attributed to discrete or molecularly driven stress memory phenomena such as the Bauschinger effect in polycrystalline metals or the Mullins effect in rubbers [39]–[41].

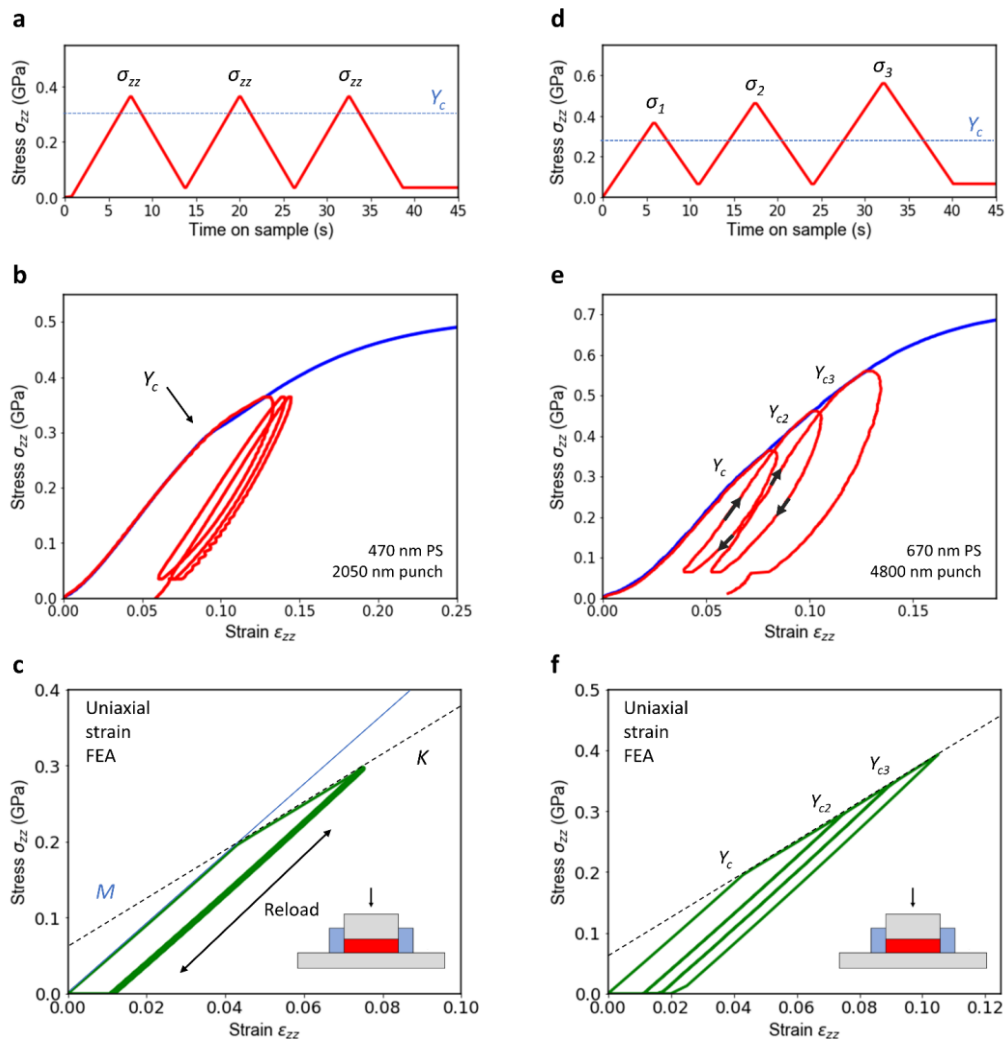

**Figure 2. Confined plastic deformation increases yield stress; cyclic loading experiments.** (a) Loading history for cyclic indentation to the same peak stress with (b) measured stress-strain curve (red) compared to simple monotonic loading curve (blue). Note the yield kink at $Y_c$ present in the initial loading, but not in the second or third loading cycle. (d) Loading history for cyclic indentation to increasing peak stress above $Y_c$ and (e) measured stress-strain curve (red) showing amplified yield stress on reload. This behaviour of the yield point is reproduced by simple-elastic perfectly-plastic finite element simulations of pure uniaxial strain using loading pattern of (a) in (c) and (d) in (f).

To explain this behaviour, we consider the following FEA simulation of ideal CC of an elastic-plastic material: An axial stress $\sigma_{zz}$ vs. strain $\varepsilon_{zz}$ curve for a single load-unload cycle shown in Fig 3a has the corresponding principal radial stress $\sigma_{rr}$ vs. $\sigma_{zz}$ curve shown in Fig 3b. During the elastic regime of loading, the relation between $\sigma_{zz}$ and $\sigma_{rr}$ is given by Eq. 1b. At yield, the slope sharply increases to $\sigma_{zz} = \sigma_{rr}$. This may be understood by considering the behaviour of the Von Mises equivalent shear stress $Q$, which in uniaxial strain geometry can be written in the form:

$$Q = \sqrt{\frac{1}{2}[(\sigma_{zz} - \sigma_{rr})^2 + (\sigma_{rr} - \sigma_{zz})^2]} = |\sigma_{zz} - \sigma_{rr}| \tag{3}$$

At yield, the material becomes incapable of supporting further shear stress of elastic origin and $Q$ becomes capped at the tensile yield stress, i.e. $Q = Y_0$. Assuming a pressure invariant yield stress, Eq. 3 implies any increase in $\sigma_{zz}$ must be balanced by an equal increase in the radial stress $\sigma_{rr}$. When the applied stress is reduced during unloading, the material recovers its elastic behaviour and the slope reverts to its initial value given by Eq. 1b. As may be seen in Fig. 3b, the difference between the elastic and plastic slopes in the $\sigma_{zz}$ versus $\sigma_{rr}$ behaviour results in a net residual radial stress upon unloading to zero applied axial stress. As was seen in Fig. 2, the effect of this residual stress on subsequent loading cycles is to increase the apparent yield stress of the indented material by increasing the path length it must travel in pressure versus shear $(P - Q)$ space to intersect the yield surface. While perhaps somewhat unexpected, we show this effect is consistent with conventional elastic plastic mechanics from the FEA simulated Q vs. P path shown in Fig. 3c. In the elastic regime, starting from zero applied load, these two quantities are related by:

$$Q = 2\left(\frac{G}{K}\right)P \tag{4}$$

where $G$ is the elastic shear modulus [17]. At yield, shear stress is capped on the Von Mises yield surface $Q = Y_0$ (dotted black line), allowing only the hydrostatic stress to increase with further loading. Unloading returns to material to elastic behaviour, with $P$ and $Q$ decreasing as $\sigma_{zz}$ is decreased. As the applied load is reduced, a point is eventually reached where the axial and radial components of stress are equal and the total net shear stress in the system is zero, even while a non-zero load remains on the punch. Further decrease of load past this point causes $Q$ to *increase*, resulting in the inflection point of Fig. 3c at $P \sim 0.18$ GPa. When the load is fully removed ($\sigma_{zz} = 0$, shown by the blue star), significant residual $P$ and $Q$ are retained within the material, which are results of the non-zero radial stress present in the material from Fig. 3b. When the material is reloaded from this point, it must retrace this path through the inflection point in $P - Q$ space to reach the yield surface a second time, which is longer than the initial path by $\sigma_p - Y_c$, where $\sigma_p$ is the peak applied stress. By this mechanism, the yield stress appears to have increased as in Fig. 2, despite no apparent change to the intrinsic structure of the material. This direct and quantifiable measure of the impact of residual stress on yield stands in contrast to previous indentation studies conducted with spherical or Berkovich tips, where the complexity of the deformation field meant that only quite limited conclusions could be drawn.[42], [43]

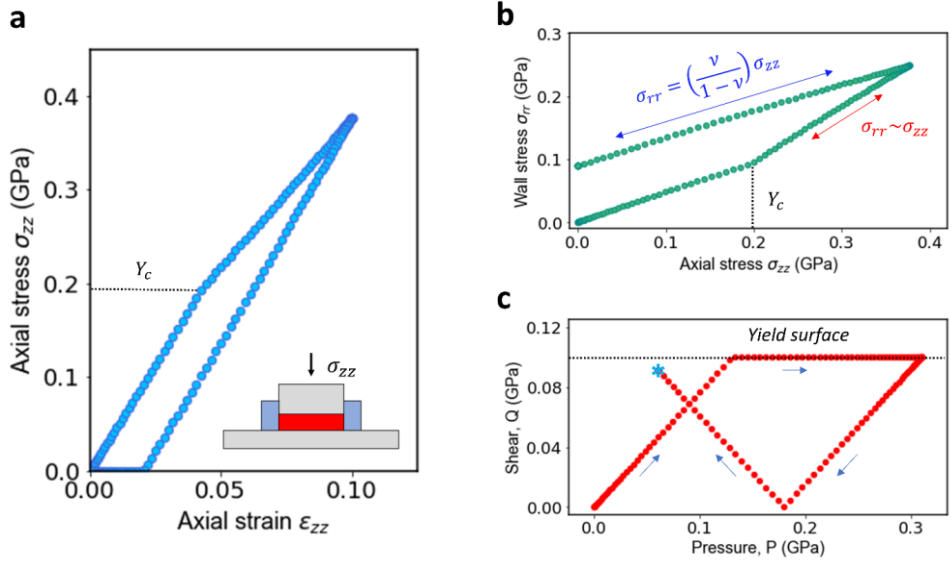

**Figure 3. Residual stresses lengthen path to yield.** (a) FEA simulated axial stress-strain curve for a simple elastic – perfectly plastic material loaded to a peak stress $\sigma_p > Y_c$ and unloaded to zero applied stress in CC. (b) Outward radial stress $\sigma_{rr}$ on the confining jacket as a function of applied stress $\sigma_{zz}$. (c) Path of the material in $P$ – $Q$ stress space during loading and unloading back to zero (blue star.)

We may further test this interpretation by considering what would happen should the confining jacket be removed, and the accumulated residual stresses be allowed to relax. This is the scenario presented in the FEA simulation of Fig 4. First the material is loaded in CC to a peak stress greater than $Y_c$. The radial confining condition is then removed, resulting in lateral expansion of the sample. Re-enforcing the confining condition and compressing the sample a second time gives the stress – strain curve shown on the right, which has been offset for clarity, where $Y_c$ is observed to revert to its initial value. Allowing the sample to expand laterally has allowed the radial and shear stresses presented in Fig. 3 (b) & (c) to relax, essentially reverting it to its pre deformation state. We note that in our experimental thin film geometry no such relaxation is possible through this mechanism as the compressed region is directly embedded in the confining jacked rendering the effect essentially permanent.

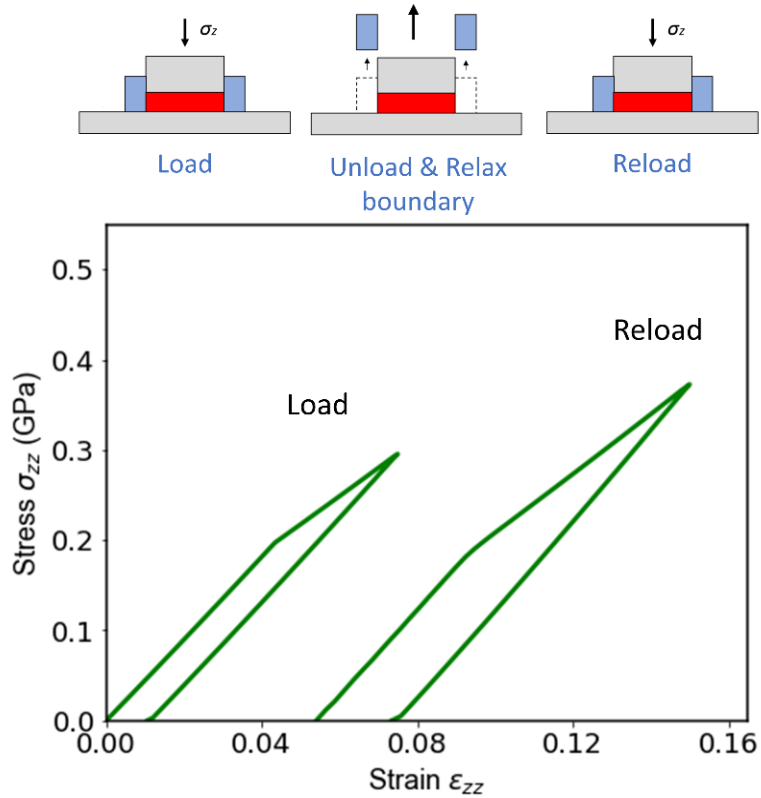

**Figure 4**. Recovery of initial yield stress upon radial relaxation.

The hardening phenomenon we describe above may in principle be experienced by any material which displays a discrete yield transition when exposed to repeated plastic deformation cycles in a confined geometry. While the mechanisms behind plastic deformation in different material classes vary greatly, e.g. dislocation slip in crystals versus shear transformation zones in glasses, so long as plastic yield results in stress – strain relations that differ from those in the elastic domain, residual stresses will be developed which will modify the subsequent yield stress. It is instructive to compare this extrinsic effect with intrinsic molecular level alterations to the material which also manifest as changes in mechanical properties. We consider two examples for polymer films; physical ageing of atactic polystyrene and thermal crosslinking of mr-I 9000, a commercially available nanoimprint micro-resist.[29], [30], [44] Addressing first the PS, physical ageing is the process through which an out of equilibrium glass tends towards a denser state through segmental rearrangement, with concurrent changes in mass density and mechanical properties such as yield stress and modulus.[45] In Fig 5 (a) we show the CC stress – strain curves for a 550 nm PS film indented with the 4800 nm punch ($\alpha = 8.7$) for two different thermal histories prior to indentation. For the quenched (blue) test, the was sample cooled rapidly from $T_g + 30°C$ to room temperature in under one minute before being mounted to the indentation stage. For the annealed (red) test, the sample was slowly cooled from $T_g + 30°C$ to $T_g - 25°C$ at a rate of 8°/hr and then held at that temperature for 5 hours before being transferred to the indenter. As ageing occurs more quickly closer to $T_g$ due to increased mobility, this thermal history produces to a denser, closer to equilibrium glass.[46] This is reflected in the stress - strain curve of Fig 5 (a), where a slightly higher yield stress is observed, as well as a higher confined plastic modulus $K$, and a higher flow stress. We note that in the extrusion limit the two curves re-intersect and follow the same path (see inset), indicating they have been driven into a new history independent state at high stresses, a possible sign of mechanically driven rejuvenation.[47]

The thermally curable micro-resist mr-I 9000 is based on a highly branched PDAP pre-polymer with an initial $T_g$ of 65°C. Heating to temperatures greater than 120°C results in thermally activated crosslinking via reaction of the allyl double bond, leading to a denser, more highly connected polymer matrix. Indentation was performed on a 300 nm film with the 2050 nm punch ($\alpha = 6.8$) before and after curing at 150°C for 30 minutes, with the CC stress – strain curves shown in Fig 5 (b). As with physical ageing in PS, curing resulted in a stiffer response in the confined plastic domain. The confined yield stress was also observed to increase from 0.32 to 0.38 GPa. We note however that there is no convergence of the curves in the extrusion limit, which we interpret as a permanent topological change made to the network via cross-linking.

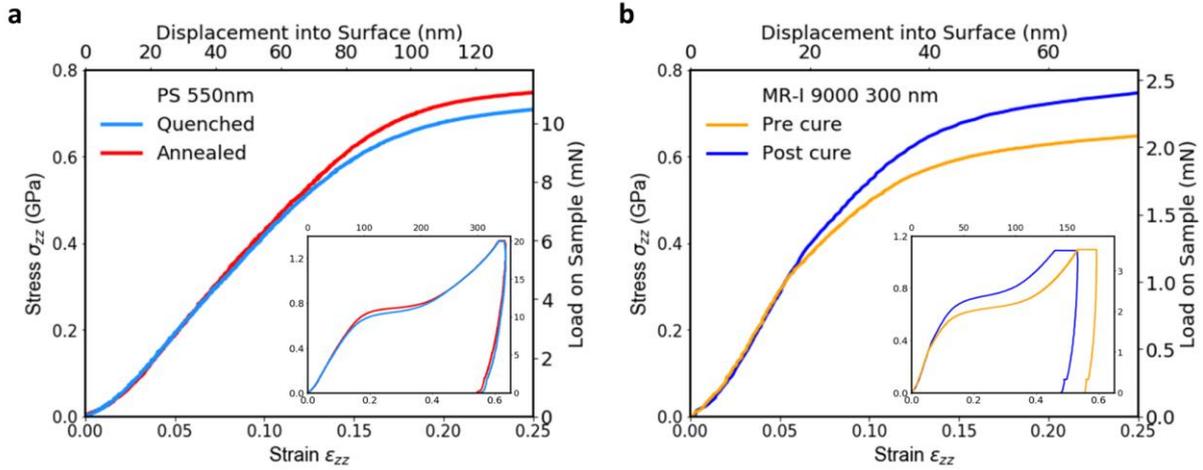

**Figure 5. Influence of microstructural changes on the stress – strain response for glassy films in confined compression.** (a) Stress – strain curves for a 550 nm PS film after quenched (blue) and annealed (red) thermal preparation histories. Inset shows full load-displacement curve, including extrusion regime. (b) Stress – strain curves for a 300 nm mr-I 9000 film before (yellow) and after (blue) thermally activated crosslinking at 150ºC.

These intrinsic changes to the microstructure of the two materials display similar mechanical signatures to our extrinsic hardening effect yet evidently have very different origins. This highlights the importance of a rigorous treatment of stresses and strains and deformation history in the analysis of any advanced material on top of consideration of the molecular reordering processes at play.

## 3. Conclusions

Confined testing geometries are becoming an increasingly common tool to elucidate the influences of pressure and shear on the molecular reordering processes associated with plastic yield.[1], [2], [18], [48] This is particularly true of non-equilibrium materials such as polymers or bulk metallic glasses, where free volume diffusion and annihilation processes are highly sensitive to the relative magnitude of these mean stresses.[49], [50] In this work we have demonstrated that one of the primary indicators of intrinsic microstructural change used in such studies, strain hardening, may be mimicked in confined compression by residual radial stresses imparted during plastic loading. A 66% increase in the indentation yield stress of atactic polystyrene was achieved via this residual strain effect, in line with the predictions of a simple linear elastic-plastic finite element model. These results and the accompanying analysis provide a useful framework to rigorously separate extrinsic and intrinsic confinement effects, and perhaps suggest that a critical reassessment of previous results in similar testing geometries may be due.

# Acknowledgements

OB and GLWC acknowledge support from Science Foundation Ireland grants SFI/AMBER/12/RC/2278 and 08/IN.1/I1932.

# Bibliography

[1]   Z. T. Wang, J. Pan, Y. Li, and C. A. Schuh, "Densification and Strain Hardening of a Metallic Glass under Tension at Room Temperature," *Phys. Rev. Lett.*, vol. 111, no. 13, p. 135504, Sep. 2013.

[2]   J. Pan, Y. P. Ivanov, W. H. Zhou, Y. Li, and A. L. Greer, "Strain-hardening and suppression of shear-banding in rejuvenated bulk metallic glass," *Nature*, vol. 578, no. 7796, pp. 559–562, Feb. 2020.

[3]   W. Xia and S. Keten, "Interfacial stiffening of polymer thin films under nanoconfinement," *Extrem. Mech. Lett.*, 2015.

[4]   J. A. Forrest, K. Dalnoki-Veress, J. R. Stevens, and J. R. Dutcher, "Effect of free surfaces on the glass transition temperature of thin polymer films," *Phys. Rev. Lett.*, vol. 77, no. 10, pp. 2002–2005, Sep. 1996.

[5]   P. A. O'Connell and G. B. McKenna, "Rheological measurements of the thermoviscoelastic response of ultrathin polymer films," *Science (80-. ).*, vol. 307, no. 5716, pp. 1760–1763, Mar. 2005.

[6]   J. M. Torres, C. M. Stafford, and B. D. Vogt, "Elastic modulus of amorphous polymer thin films: Relationship to the glass transition temperature," *ACS Nano*, vol. 3, no. 9, pp. 2677–2685, Sep. 2009.

[7]   H. D. Rowland, W. P. King, J. B. Pethica, and G. L. W. Cross, "Molecular Confinement Accelerates Deformation of Entangled Polymers During Squeeze Flow," *Science*, no. October, pp. 720–724, 2008.

[8]   H. D. Rowland, W. P. King, G. L. W. Cross, and J. B. Pethica, "Measuring Glassy and Viscoelastic Polymer Flow in Molecular-Scale Gaps Using a Flat Punch Mechanical Probe," *ACS Nano*, vol. 2, no. 3, pp. 419–428, Mar. 2008.

[9]   W. Guo *et al.*, "Intrinsic and extrinsic size effects in the deformation of amorphous CuZr/nanocrystalline Cu nanolaminates," *Acta Mater.*, vol. 80, pp. 94–106, Nov. 2014.

[10]  D. R. P. Singh, N. Chawla, G. Tang, and Y. L. Shen, "Micropillar compression of Al/SiC nanolaminates," *Acta Mater.*, vol. 58, no. 20, pp. 6628–6636, Dec. 2010.

[11]  G. B. McKenna, "Size and confinement effects in glass forming liquids : Perspectives on bulk and nano-scale behaviours," *Le J. Phys. IV*, vol. 10, no. PR7, pp. Pr7-53-Pr7-57, May 2000.

[12]  J. R. Greer and J. T. M. De Hosson, "Plasticity in small-sized metallic systems: Intrinsic versus extrinsic size effect," *Prog. Mater. Sci.*, vol. 56, no. 6, pp. 654–724, Aug. 2011.

[13]  G. Dieter, *Mechanical Metallurgy*, 3rd ed. New York: McGraw-Hill, 1986.

[14]  D. Tabor, *The Hardness of Metals*. Oxford: Clarendon Press, 1951.

[15]  K. L. Johnson, "The correlation of indentation experiments," *J. Mech. Phys. Solids*, vol. 18, no. 2, pp. 115–126, 1970.

[16]  K. L. Johnson, *Contact Mechanics*. Cambridge: Cambridge University Press, 1985.

[17]  O. Brazil *et al.*, "In situ measurement of bulk modulus and yield response of glassy thin films via confined layer compression," *J. Mater. Res.*, vol. 35, no. 6, pp. 644–653, Mar. 2020.

[18]  K. Ravi-Chandar and Z. Ma, "Inelastic Deformation in Polymers under Multiaxial Compression," *Mech. Time-Dependent Mater.*, vol. 4, no. 4, pp. 333–357, 2000.

[19]  M. J. Adams, M. A. Mullier, and J. P. K. Seville, "Agglomerate strength measurement using a uniaxial confined compression test," *Powder Technol.*, vol. 78, no. 1, pp. 5–13, Jan. 1994.

[20]  G. Mavko, T. Mukerji, and J. Dvorkin, *The Rock Physics Handbook*. Cambridge: Cambridge University Press, 2009.

[21]  O. Brazil, "Deformation and Yield of Polymer Thin Films Under Confinement," 2018.

[22]  P. M. Anderson and C. Li, "Hall-Petch relations for multilayered materials," *Nanostructured Mater.*, vol. 5, no. 3, pp. 349–362, Mar. 1995.

[23]  E. Arzt, "Size effects in materials due to microstructural and dimensional constraints: a comparative review," *Acta Mater.*, vol. 46, no. 16, pp. 5611–5626, Oct. 1998.

[24]  J. W. Kim, G. A. Medvedev, and J. M. Caruthers, "Observation of yield in triaxial deformation of glassy polymers," *Polymer (Guildf).*, vol. 54, no. 11, pp. 2821–2833, 2013.

[25]  J. W. Kim, G. A. Medvedev, and J. M. Caruthers, "Mobility evolution during tri-axial deformation of a glassy polymer," *Polymer (Guildf).*, vol. 55, no. 6, pp. 1570–1573, 2014.

[26]  J. Pan, Y. X. Wang, Q. Guo, D. Zhang, A. L. Greer, and Y. Li, "Extreme rejuvenation and softening in a bulk metallic glass," *Nat. Commun.*, vol. 9, no. 1, p. 560, Dec. 2018.

[27]  J. Pan, Y. X. Wang, and Y. Li, "Ductile fracture in notched bulk metallic glasses," *Acta Mater.*, vol. 136, pp. 126–133, Sep. 2017.

[28]  J. Pan, H. F. Zhou, Z. T. Wang, Y. Li, and H. J. Gao, "Origin of anomalous inverse notch effect in bulk metallic glasses," *J. Mech. Phys. Solids*, vol. 84, pp. 85–94, Aug. 2015.

[29]  H. Schulz *et al.*, "New polymer materials for nanoimprinting," *J. Vac. Sci. Technol. B Microelectron. Nanom. Struct.*, vol. 18, no. 4, pp. 1861–1865, Aug. 2000.

[30]  K. Pfeiffer *et al.*, "A comparison of thermally and photochemically cross-linked polymers for nanoimprinting," in *Microelectronic Engineering*, 2003, vol. 67–68, pp. 266–273.

[31]  J. Brandrup, E. H. Immergut, and E. A. Grulke, *Polymer handbook*. Wiley-Interscience, 1999.

[32]  M. D. Ediger, H. N. Lee, K. Paeng, and S. F. Swallen, "Dye reorientation as a probe of stress-induced mobility in polymer glasses," *J. Chem. Phys.*, vol. 128, no. 13, 2008.

[33]  D. J. Lacks and M. J. Osborne, "Energy landscape picture of overaging and rejuvenation in a sheared glass," *Phys. Rev. Lett.*, vol. 93, no. 25, pp. 1–4, 2004.

[34] M. L. Wallace and B. Joós, "Shear-Induced Overaging in a Polymer Glass," *Phys. Rev. Lett.*, vol. 96, no. 2, p. 025501, Jan. 2006.

[35] M. L. Falk and J. S. Langer, "Deformation and Failure of Amorphous, Solidlike Materials," *Annu. Rev. Condens. Matter Phys.*, vol. 2, no. 1, pp. 353–373, Mar. 2011.

[36] E. F. Oleinik, M. A. Mazo, M. I. Kotelyanskii, S. N. Rudnev, and O. B. Salamatina, "Plastic deformation in disordered solids: The state of the art and unresolved problems," in *Advanced Structured Materials*, vol. 94, Springer Verlag, 2019, pp. 313–332.

[37] M. L. Falk and J. S. Langer, "Dynamics of viscoplastic deformation in amorphous solids," *Phys. Rev. E - Stat. Physics, Plasmas, Fluids, Relat. Interdiscip. Top.*, vol. 57, no. 6, pp. 7192–7205, Jun. 1998.

[38] P. Schall, D. A. Weitz, and F. Spaepen, "Structural rearrangements that govern flow in colloidal glasses," *Science (80-. ).*, vol. 318, no. 5858, pp. 1895–1899, Dec. 2007.

[39] Y. Xiang and J. J. Vlassak, "Bauschinger effect in thin metal films," *Scr. Mater.*, vol. 53, no. 2, pp. 177–182, Jul. 2005.

[40] J. Diani, B. Fayolle, and P. Gilormini, "A review on the Mullins effect," *European Polymer Journal*, vol. 45, no. 3. Pergamon, pp. 601–612, 01-Mar-2009.

[41] S. Patinet, A. Barbot, M. Lerbinger, D. Vandembroucq, and A. Lemaître, "Origin of the Bauschinger Effect in Amorphous Solids," *Phys. Rev. Lett.*, vol. 124, no. 20, p. 205503, May 2020.

[42] G. M. Pharr, T. Y. Tsui, A. Bolshakov, and W. C. Oliver, "Effects of residual stress on the measurement of hardness and elastic modulus using nanoindentation," in *Materials Research Society Symposium - Proceedings*, 1994, vol. 338, pp. 127–134.

[43] J. G. Swadener, B. Taljat, and G. M. Pharr, "Measurement of residual stress by load and depth sensing indentation with spherical indenters," *J. Mater. Res.*, vol. 16, no. 7, pp. 2091–2102, 2001.

[44] K. Pfeiffer *et al.*, "Novel linear and crosslinking polymers for nanoimprinting with high etch resistance," *Microelectron. Eng.*, vol. 53, no. 1, pp. 411–414, Jun. 2000.

[45] C. B. Roth, *Polymer glasses*, 1st ed. Boca Raton: CRC Press, 2016.

[46] S. L. Simon, J. W. Sobieski, and D. J. Plazek, "Volume and enthalpy recovery of polystyrene," *Polymer (Guildf).*, vol. 42, no. 6, pp. 2555–2567, Mar. 2001.

[47] G. B. McKenna, "Mechanical rejuvenation in polymer glasses: fact or fallacy ?," *J. Phys. Condens. Matter*, vol. 15, p. S737, 2003.

[48] Z. Ma and K. Ravi-Chandar, "Confined compression: A stable homogeneous deformation for constitutive characterization," *Exp. Mech.*, vol. 40, no. 1, pp. 38–45, Mar. 2000.

[49] F. Spaepen, "A microscopic mechanism for steady state inhomogeneous flow in metallic glasses," *Acta Metall.*, vol. 25, no. 4, pp. 407–415, 1977.

[50] F. Spaepen, "Homogeneous flow of metallic glasses: A free volume perspective," *Scr. Mater.*, vol. 54, no. 3, pp. 363–367, Feb. 2006.